\documentclass[12pt]{extarticle}
\usepackage[mathscr]{euscript}
\usepackage{amsmath}
\usepackage[numbers]{natbib}
\usepackage{graphicx}
\usepackage{xcolor}

\usepackage[numbers]{natbib}
\usepackage{graphicx}
\usepackage{hyperref}

\title{US faces endemic Covid-19 infections and deaths; ways to stop the pandemic}

\author{Fazle Hussain$^*$,\\
Zeina S. Khan,\\
\&\\
Frank Van Bussel\\
Texas Tech University, Department of Mechanical Engineering\\ 
2703 7th Street, Box: 41021, Lubbock, TX 79409\\
Phone: 832-863-8364\\ 
$*$ fazle.hussain@ttu.edu\\}

\begin{document}
\maketitle
\begin{abstract}
A new epidemic model for Covid-19 has been constructed and simulated for eight US states. The coefficients for this model, based on seven coupled differential equations, are carefully evaluated against recorded data on cases and deaths. These projections reveal that Covid-19 will become endemic, spreading for more than two years. If stay-at-home orders are relaxed, most states may experience a secondary peak in 2021. The number of Covid-19 deaths could have been significantly lower in most states that opened up, if lockdowns had been maintained. Additionally, our model predicts that decreasing contact rate by 10\%, or increasing testing by approximately 15\%, or doubling lockdown compliance (from the current  $\sim$ 15\%) will eradicate infections in Texas within a year. Applied to the entire US, the predictions based on the current situation indicate about 11 million total infections (including undetected), 8 million cumulative confirmed cases, and 630,000 cumulative deaths by November 1, 2020.

\end{abstract}
The first cases of community Covid-19 transmission in the United States were reported in California, Oregon, Washington state, and New York state in late February, 2020  -- roughly two months after the initial outbreak in Wuhan, China \cite{schu2020how}. A national emergency was declared by the US President on March 13, 2020, and testing several days later revealed that Covid-19 had spread to all 50 states. On March 20, New York City was declared the US outbreak epicenter \cite{schu2020how}. As of June 29, 2020, the US had approximately 2.5 million confirmed Covid-19 cases and 125,000 confirmed deaths \cite{world2020coronavirus161}; we note that Florida, Arizona, and Texas are now emerging outbreak epicenters. It is natural to ask how the situation will look in a few months, and even in a few years, based on the recent trend of rapidly rising cases in several US states, and also to ask what interventions need to be ramped up in order to prevent further spread.  
\begin{figure}[htb!]
\centering
\includegraphics[width=0.85\linewidth]{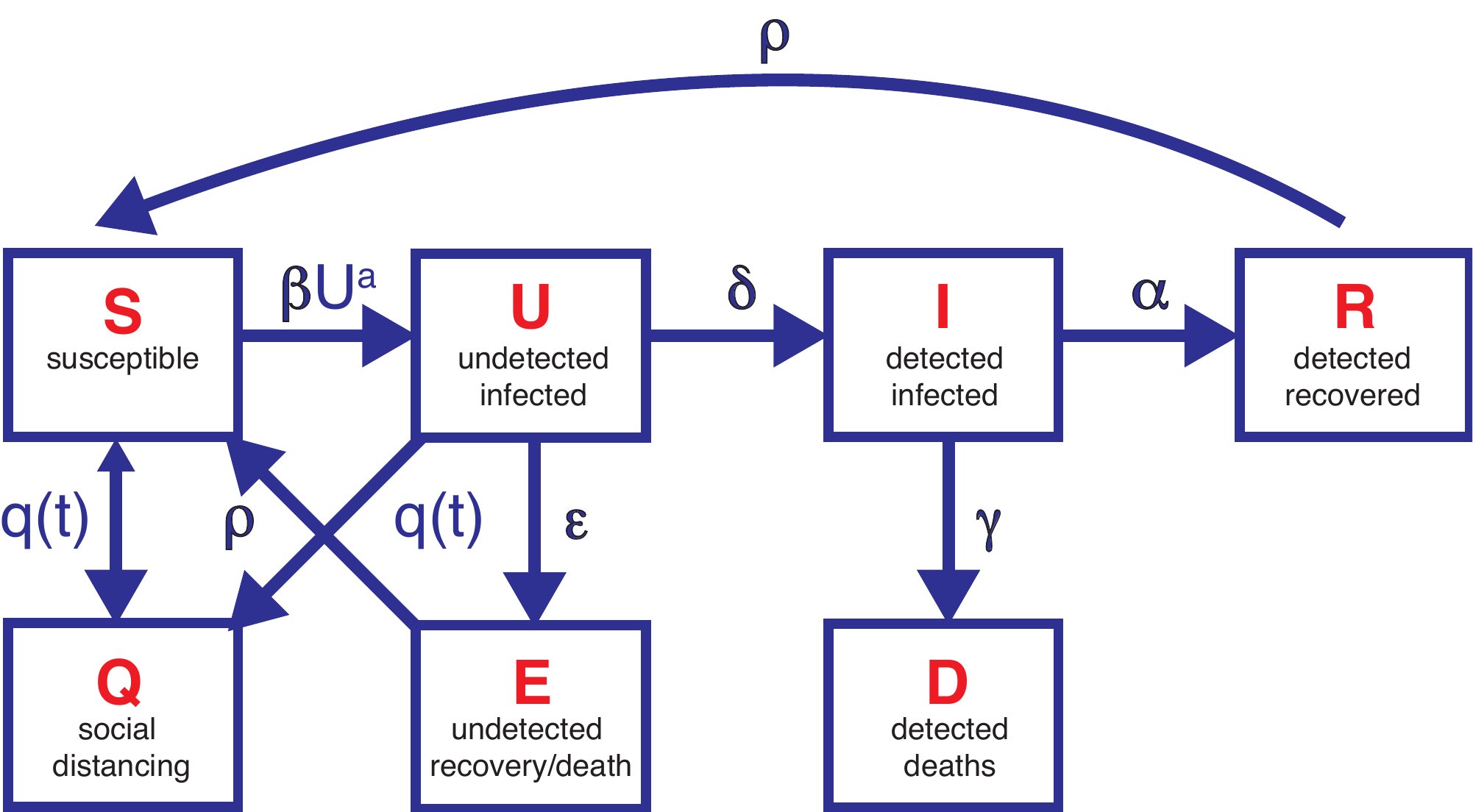}
\caption{A new epidemiological compartment model for Covid-19. Each compartment reflects a category of the population, with rates of transfer between compartments indicated by the arrows. $\beta U^a$  is the infection rate, $q(t)$ is the sequestration rate, $\epsilon$ is the exclusion rate due to undetected recovery or death, $\delta$ is the detection rate, $\alpha$ is the detected recovery rate, $\gamma$ is the detected death rate, and $\rho$ is the reintroduction rate (due to loss of immunity as reinfection is quite likely, particularly for Covid-19 \cite{long2020clinical} -- either due to intrinsic loss of antibody effectiveness or due to the virus having metamorphosed). 		
}
\label{Ffig1}
\end{figure}

We propose a new compartmental, epidemiological, dynamical systems model (see \cite{brauer2019simple}) to predict Covid-19 virus growth. It considers: both detected and undetected infected populations, medical quarantine and social sequestration, release from sequestration, plus possible reinfection due to loss of immunity. The model considers the population to be conceptually separated into mutually exclusive compartments or subgroups: susceptible (which is the majority of the population at the beginning of an outbreak), undetected infecteds (who presumably avoid detection due to having mild or no symptoms), excluded (due to undetected recovery or deaths), quarantine (not representative of an enforced medical quarantine -- which is already accounted for by the detected infected compartment -- but rather self-sequestration), detected infected (detected due to testing), detected recovered, and detected deaths. Individuals are transferred between compartments at distinct rates, as shown in Figure \ref{Ffig1}, where the rates are determined by fitting to reliable and verified data on case counts and deaths. For our study, the data was obtained from the Johns Hopkins University's GitHub website repository \cite{csse2020data}. Sheltering-at-home, as some states have ordered, and release from this, since some states have reopened, are modeled by having a time dependent sequestration rate $q(t)$ with two peaks centered on the days these measures impacted each state -- the days and the magnitude of the peaks (corresponding to the fraction of the population transferred) are also fit parameters.

Note that the essential process behind our study is simple. First we write a set of seven coupled ordinary differential equations, some nonlinear, for the compartment content level; these equations contain a series of coefficients, one is time-dependent, which denote the transfer rates between the compartments. These coefficients are evaluated by fitting the model equations to data, and then the evolutionary solutions of the equations for the compartment levels can be simply calculated using a computer. Sensitivity to these coefficients can be explored by altering their values and studying the solutions. 

As is usual with modeling complex phenomena such as Covid-19, a major challenge is to focus on the most important aspects of the phenomenon by undertaking judicious simplifications. Hence, in our study, several simplifying assumptions or idealizations have been made. We assume that detected infected individuals do not transmit Covid-19 to the susceptible population during medical or self-isolation. Our contact rate $\beta$, and therefore the infectious power $\beta U^a$ (see Figure \ref{Ffig1}), is modeled as being constant in time -- in future versions of our model we may incorporate time dependent $\beta$ or $a$ in order to disentangle population wide transmission suppression (e.g. face masks) from social contact suppression (stay-at-home orders). Our $q(t)$ function removes susceptibles and undetetected infecteds at the same rate (we have no reason at this time to differentiate the rates of change of these populations). Similarly, people from the excluded and recovered compartments lose immunity at the same rate $\rho$. Indeed, it is possible that people who experienced milder forms of the disease (excluded) lose immunity faster than those who experienced more severe and persistent symptoms and sought out treatment (recovered); this is not unexpected because severe illness is quite likely to induce higher levels of antibodies \cite{long2020clinical}; however, we use a single rate for simplicity. Finally, we do not consider the effects of births, vertical transmissions (from mother to child during pregnancy), immigrants, emigrants, or deaths due to other diseases or trauma. 

\begin{figure}[htb!]
\centering
\includegraphics[width=0.85\linewidth]{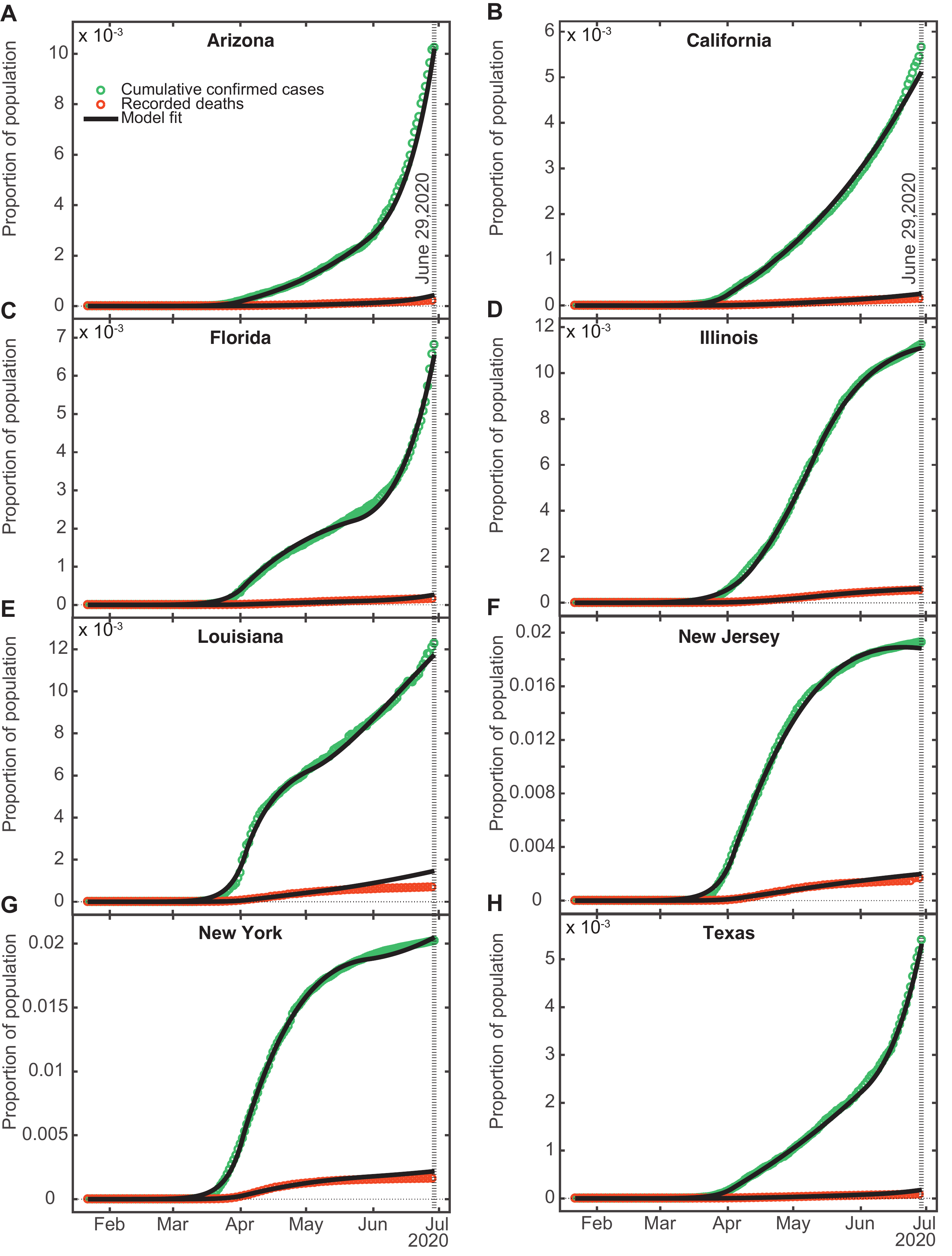}
\caption{Fits of our new compartment model to confirmed cumulative case counts and deaths for (A) Arizona, (B) California, (C) Florida, (D) Illinois, (E) Louisiana, (F) New Jersey, (G) New York state, and (H) Texas. Reliable data was obtained from The Johns Hopkins University \cite{csse2020data}. The vertical dashed line indicates the last date fitting data was obtained for -- June 29, 2020.}
\label{Ffig2}
\end{figure}

The coefficients in the model, including contact rate $\beta$, $U(0)$ at the beginning of the outbreak (time 0), $a$, and detection $\delta$, recovery $\alpha$, sequestration and reintroduction (into the susceptible class) $q(t)$ rates,  are evaluated by fitting to empirical case and death data for eight major US states: Arizona, California, Florida, Illinois, Louisiana, New Jersey, New York State, and Texas -- see figure \ref{Ffig2}. Together, these states make up 43\% of the US population; some of these states appear to have handled their initial outbreaks well, while others appear to be emerging hotspots. The evolution of Covid-19 is fairly similar among these states: variations in contact and recovery rates remain below 5\%; however, variations are larger in death rate, reinfection rate, sequestration rate, and release rate from sequestration. Surprisingly, we find that the sequestration rate is low, less than 20\%, in all states considered. The release rate from sequestration is rather high in some states -- around 50\% in Arizona and Florida, and nearly 40\% in Texas, explaining the current dire situation of rapidly rising case counts.

Extending our model based on the current situation out to two years, we find that Covid-19 will become endemic, spreading for more than two years, with the peak number of cases occurring for most states in the Autumn months -- see Figure 3 (gray dashed curve). Had lockdowns been continued at the prior compliance rate, the number of cases would have been one tenth, and death rates up to  five times  lower. Fitting our model to all of the US states and generating predictions based on the current situation, we find that there will be about 11 million total infections (including undetected), 8 million cumulative confirmed cases, and 630,000 cumulative deaths by November 1, 2020. 

\begin{figure}[htb!]
\centering
\includegraphics[width=0.85\linewidth]{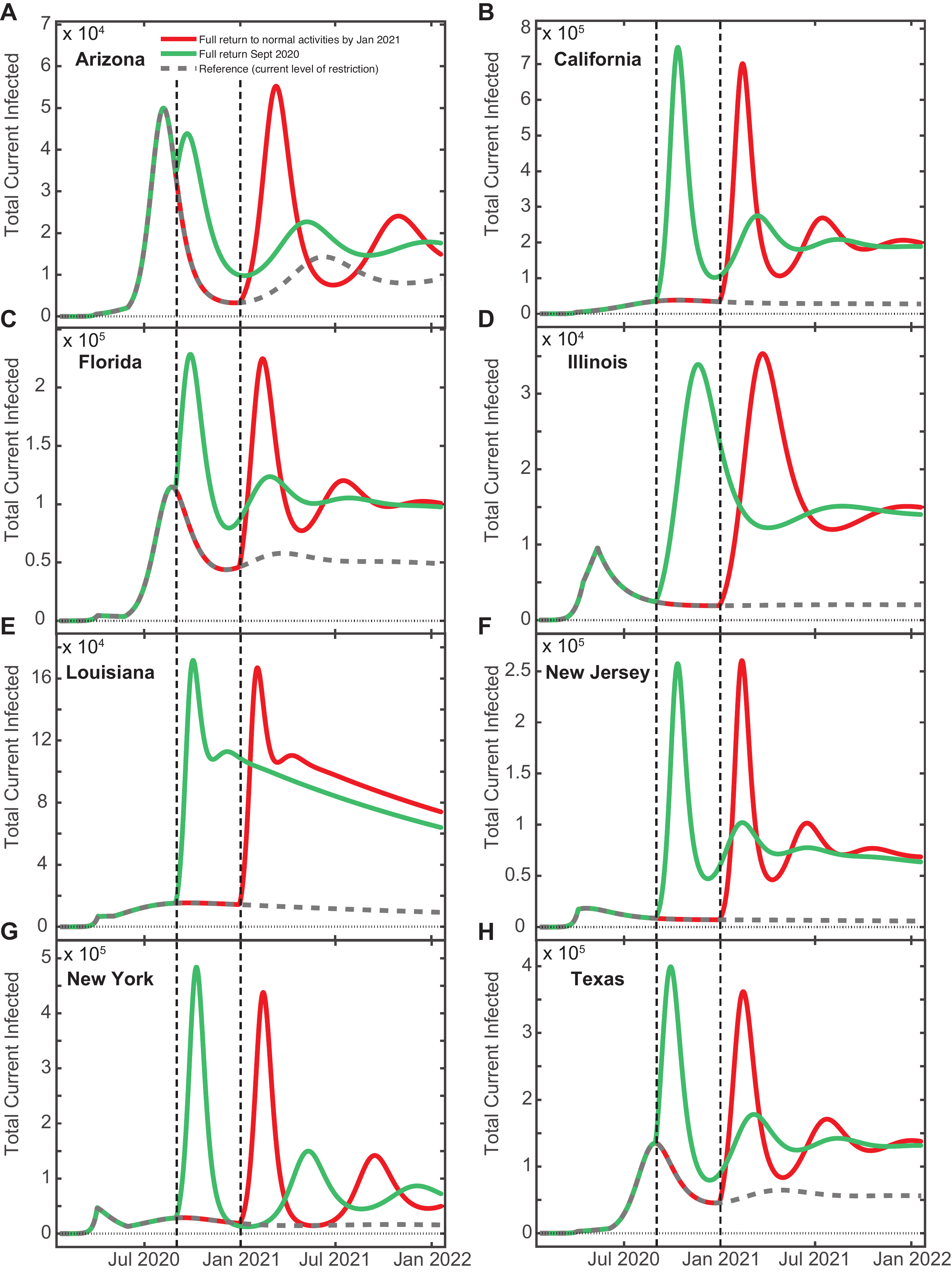}
\caption{Consequences of total reopening for different states. Total infected, for September 1, 2020 release (green curve) and January 1, 2021 (red curve), shown alongside our current forecast (gray dashed line) for: (A) Arizona, (B) California, (C) Florida, (D) Illinois, (E) Louisiana, (F) New Jersey, (G) New York, and (H) Texas. Vertical dashed lines delineate the day of lockdown release. Note that the two spikes are nearly equally strong -- as a consequence of having excessive active cases at the time of reopening -- and hence the final values essentially converge.}
\label{Ffig3}
\end{figure}

Considering that some states are planning for a phased reopening, with further access to businesses and entertainment venues in the coming months, we examined the effects of fully reopening all of the states considered on two dates -- September 1, 2020, and January 1, 2021 -- see Figure \ref{Ffig3}. For fully reopening on both of these dates, we observe a similar trend -- the number of infections rapidly rise, leading to peaks in infections within two months of reopening -- similar to what is currently occurring in Arizona, Florida, and Texas. We also observe that many states experience a smaller, second peak in infections within six months of the first peak. This demonstrates that reopening states should not occur unless additional interventions to reduce the infectiousness of the disease and its spread are implemented.

Based on the current situation in Texas, we tested the effects of interventions on Covid-19 prevalence. We found that decreasing the contact rate by only 10\% results in eradication of the disease within one year. Such an increase is easily achievable by people in the absence of formal statewide orders simply by wearing masks and increasing social distancing. Increasing detection by testing by 15\% also results in eliminating Covid-19 within a year, as does doubling the current compliance with staying at home (from the current 15\% to 30\%). These findings are somewhat surprising and suggest that it is indeed possible to eliminate the virus, which will otherwise remain endemic. 

Please find our recent preprint here: \url{https://arxiv.org/abs/2006.05955?fbclid=IwAR1ESlf4mzJFkHJLt\_iGCE0OTtYgSVLPt-iVgPUtg4JU7ylQscGCTBjFUNA}.


\end{document}